\documentstyle[aps]{revtex}

\begin{document}
\title{Exact solutions of nonlinear PDE, nonlinear transformations and reduction
nonlinear PDE to a quadrature}
\author{Yang Lei, Liu Jianbin, Yang Kongqing}
\address{{\it Department of physics, Lanzhou University, Gansu 730000, China}}
\maketitle

\begin{abstract}
A method to construct the exact solution of the PDE is presents, which
combines the two kind methods(the nonlinear transformation and RQ(Reduction
the PDE to a Quadrature problem) method).The nonlinear diffusion equation is
chosen to illustrate the method and the exact solutions are obtained.
\end{abstract}

Seeking the exact solutions of the nonlinear PDE(Partial Differential
Equation) play an important role in the nonlinear problems, a number of
methods have been developed, such as, Inverse Scattering theory[1], but in
some cases[2,3], these methods cannot do very well. So the simple and direct
methods to find analytic solutions of PDE have drawn a lot of interest, for
example, Hirota's bilinear methods[4], the truncated Painlev\.{e}
expansion[5,6], homogeneous balance method[7,8], reduction the PDE to a
quadrature problem[9], the hyperbolic tangent function-series method[10] and 
{\it etc}.

In the sense of the seeking solutions, these direct methods can be divided
into two kinds, one kind bases on the nonlinear transformation, such as,
Hirota's bilinear methods bases on the Hirota transformation, by using the
transformation the nonlinear PDE becomes the bilinear forms, then seeking
the solution of the bilinear forms leads to obtain the solution of the
origin PDE. The truncated Painlev\.{e} expansion bases on the Painlev\.{e}
expansion, by using the leading-order analysis a set of
Painlev\.{e}-B\"{a}cklund equations are obtained, then seeking the solution
of the set of Painlev\.{e}-B\"{a}cklund equations leads to obtain the
solution of the origin PDE. The homogeneous balance method gives the
nonlinear transformation based on a simple program, by using the
transformation a set of equations are obtained, then seeking the solution of
the set of equations leads to obtain the solution of the origin PDE. And the
special nonlinear transformation can be given for the special nonlinear PDE.

Another kind of method bases on the direct ansatz of the solution, most of
those methods are applied to seek the travelling wave solution, such as, the
hyperbolic tangent function-series method and Reduction the PDE to a
Quadrature problem(RQ method). In generalized, the nonlinear PDE\ is written
as 
\begin{equation}
P(u,u_x,u_t,u_{xx},u_{xt},u_{tt},\cdots \cdots )=0.  \label{1}
\end{equation}
Introducing the similarity variable $\xi =x-vt$, the travelling wave
solutions $u(\xi )=u(x-vt)$ satisfy

\begin{equation}
P^{\prime }(u,u_\xi ,u_{\xi \xi },\cdots \cdots )=0.  \label{2}
\end{equation}
The hyperbolic tangent function-series method makes the ansatz,

\begin{equation}
u(\xi )=\sum\limits_{i=0}^ma_i(tanh(b\xi ))^i,  \label{3}
\end{equation}
where the velocity $v$, the integer $m$, the real $a_i$ and $b$ are
parameter to be determined, inserting the ansatz into the equation(2), a set
of the nonlinear algebra equations of these undetermined parameters are
obtained, solving it, the solution of the equation(1) is obtained. Reduction
the PDE to a Quadrature problem(RQ method) is an interesting method for
looking for the travelling wave solutions, the essence of the method can be
presented in this way: First, giving a ansatz for $\frac d{dx}u(x,t)$,
namely,

\begin{equation}
u_\xi =\sum_{i=0}^Na_iu^{b_i}(\xi ),  \label{4}
\end{equation}
where $a_i$, $b_i$ are undetermined constants(usually $N\leq 3$),
substituting the formula (4) into the equation (2), when the suitable
constants $a_i$, $b_i$ are chosen, the equation (2) becomes a identical
equation. Then solving the equation (2) becomes to solve the equation (4),
it's obviously the solution of the equation (4) can be obtained just by
integration.

This paper try to combine the two kinds of methods to solve the nonlinear
PDE. The first step, the origin PDE is transformed into other equations by a
nonlinear transformation, then RQ method is applied to these equations to
obtain the exact solution. Here 1+1 D nonlinear diffusion equation is
chosen, 
\begin{equation}
u_t-u_{xx}+P(u)=0,  \label{5}
\end{equation}
to illustrate this method, where $u$ is a real scalar field defined over the
spatial $x$ and temporal $t$, $P(u)=\sum\limits_{i=0}^da_iu^i$ is a
polynomial of $u$, for most of physical problems require $d\leq 5$, we choose

\begin{equation}
P(u)=\sum\limits_{i=0}^da_iu^i=a_0+a_1u+a_2u^2+a_3u^3+a_4u^4+a_5u^5,
\label{6}
\end{equation}
where $a_i$ are parameters. When $a_0=a_3=a_4=a_5=0$, the equation(5)
becomes the Fisher equation. When $a_0=a_2=a_4=a_5=0$, the equation(5)
becomes the Newell-Whitehead equation (also known as the
Kolmogorov-Petrovsky-Piskunov equation). This paper is organized as follows.
In Sec II, the fisher equation is discussed, combining the truncated
Painlev\.{e} expansion and the RQ method, the exact solution of the fisher
equation is obtained. In Sec III, the equation $%
u_t-u_{xx}+a_1u+a_2u^2+a_3u^3+a_4u^4=0$ and equation $%
u_t-u_{xx}+a_0+a_1u+a_4u^4=0$ are discussed, combines a special
transformation for the equation and the RQ method, the exact solutions of
these equations are obtained. In Sec IV, the fifth order the NW(KPP)
equation is discussed, combines a special transformation for the NW equation
and the RQ method, the exact solution of the NW equation is obtained. In Sec
V, the main conclusions are listed.

1. In this section, the Fisher equation is discussed, which is

\begin{equation}
u_t-u_{xx}+au+bu^2=0.  \label{7}
\end{equation}
Taking the truncated Painlev\'{e} expansion,

\begin{equation}
u(x,t)=f(x,t)^{-I}\sum\limits_{l=0}^Iu(x,t)_lf(x,t)^l,  \label{8}
\end{equation}
where $I$ is a natural number to be determined, $u(x,t)_l$ is analytic
functions. The leading-order analysis for the equation(7) yields $I=2$, so
that

\begin{equation}
u(x,t)=\frac{u_2(x,t)}{f(x,t)^2}+\frac{u_1(x,t)}{f(x,t)}+u_0(x,t).  \label{9}
\end{equation}
Substituting (9) into (7) with {\it Mathematica}, making the coefficients of
like powers of vanish, so as to obtain the following set of
Painlev\'{e}-B\"{a}cklund equations,

\begin{equation}
\begin{array}{l}
f^{-1}:au_1+2bu_0u_1+u_{1t}-u_{1xx}=0; \\ 
f^{-2}:bu_1^2+au_2+2bu_0u_2-u_1f_t+u_{2t}+2f_xu_{1x}+u_1f_{xx}-u_{2xx}=0; \\ 
f^{-3}:bu_1u_2-u_2f_t-u_1f_x^2+2f_xu_{2x}+u_2f_{xx}=0; \\ 
f^{-4}:bu_2-6f_x^2=0.
\end{array}
\label{10}
\end{equation}
Here $u_0(x,t)$ satisfied

\begin{equation}
u_{0t}-u_{0xx}+a_1u_0+a_2u_0^2=0,  \label{11}
\end{equation}
which has the similar from of the equation (7), thus the set of the
equations (9)-(11) constitutes an auto-B\"{a}cklund transformation. It's
obviously, the equations(11) have a trivial solution $u_0(x,t)=0$. Based on
the trivial solution, introducing the similarity variable $\xi =x-vt$, and
taking the following ansatz

\begin{equation}
f^{\prime }(\xi )=\alpha f(\xi )+\beta ,  \label{12}
\end{equation}
where $\alpha $ and $\beta $ are undetermined parameters. Substituting (12)
into the equations (10), choosing $a=6\alpha ^2$, $v=-5\alpha $, the
equations (10) become identical equations, then the equation(12) can be
solved just by integration. So the solution of the equation (7) is

\begin{equation}
u(x,t)=-\frac{6ce^{\alpha x}\alpha ^3(c\alpha e^{\alpha x}-2\beta e^{5\alpha
^2t})}{b(c\alpha e^{\alpha x}-\beta e^{5\alpha ^2t})}.  \label{13}
\end{equation}

2. In this section, the equation $u_t-u_{xx}+a_1u+a_2u^2+a_3u^3+a_4u^4=0$
and equation $u_t-u_{xx}+a_0+a_1u+a_4u^4=0$ are discussed. The first
equation is given

\begin{equation}
u_t-u_{xx}+au+bu^2+cu^3+du^4=0.  \label{14}
\end{equation}
A transformation for equation(14) is introduced

\begin{equation}
u(x,t)=\frac{g(x,t)}{f(x,t)^{\frac 23}}-u_0(x,t),  \label{15}
\end{equation}
substituting the formula(15) into the equation(14) with {\it Mathematica},
make the coefficients of like powers of $f(x,t)$ vanish, so as to obtain the
following set equations,

\begin{equation}
\begin{array}{l}
f^{-\frac 23}:ag-2bgu_0+3cgu_0^2-4dgu_0^3+g_t-g_{xx}=0, \\ 
f^{-\frac 43}:b-3cu_0+6du_0^2=0; \\ 
f^{-\frac 53}:gf_t-2g_xf_x-gf_{xx}=0; \\ 
f^{-\frac 63}:c-4du_0=0; \\ 
f^{-\frac 83}:9dg^3-10f_x^2=0
\end{array}
\label{16}
\end{equation}
and $u_0(x,t)$ satisfied the equation

\begin{equation}
u_{0t}-u_{0xx}+au_0+bu_0^2+cu_0^3+du_0^4=0,  \label{17}
\end{equation}
which has the similar from of the equation (14), thus the set of the
equations (15)-(17) constitutes an auto-B\"{a}cklund transformation. It's
obviously, the equations(17) have a trivial solution $u_0(x,t)=0$. Based on
the trivial solution, introducing the similarity variable $\xi =x-vt$, and
taking the following ansatz 
\begin{equation}
f^{\prime }(\xi )=\alpha f(\xi )+\beta f(\xi )^2,  \label{18}
\end{equation}
where $\alpha $ and $\beta $ are undetermined parameters. Substituting (18)
into the equations (16), choosing $\alpha =\frac{3c^{\frac 32}}{8\sqrt{10}d}$%
, $b=\frac{3c^2}{8d}$, $a=\frac{3c^3}{64d^2}$, $v=-\frac{7\alpha }3$. The
equations (16) become identical equations, then the equation(18) can be
solved just by integration. So the solution of the equation (14) is

\begin{equation}
u=\frac{4\left( \frac{30}d\right) ^{\frac 13}d\left( \frac{\alpha
^2e^{\alpha x+\alpha vt+c_1}}{\left( e^{c_1}-e^{\alpha \left( x+vt\right)
}\beta \right) ^2}\right) ^{\frac 23}-3c\left( \frac{\alpha e^{\alpha (x+vt)}%
}{e^{c_1}-e^{\alpha (x+vt)}\beta }\right) ^{\frac 23}}{12d\left( \frac{%
\alpha e^{\alpha (x+vt)}}{e^{c_1}-e^{\alpha (x+vt)}\beta }\right) ^{\frac 23}%
}.  \label{19}
\end{equation}

The second equation is

\begin{equation}
u_t-u_{xx}+au+du^4=0,  \label{20}
\end{equation}
the transformation (15) is workable for the equation(20). The
auto-B\"{a}cklund transformation can be obtained similarly. Based on the
trivial solution $u_0(x,t)=0$, introducing the similarity variable $\xi =x-vt
$, and taking the following ansatz 
\begin{equation}
f^{\prime }(\xi )=\alpha +\beta f(\xi ),  \label{21}
\end{equation}
where $\alpha $ and $\beta $ are undetermined parameters. Choosing $a=-\frac{%
10\beta ^2}9$, $v=\frac{7\beta }3$, then the solution of the equation (20) is

\begin{equation}
u=\frac{10}d^{\frac 13}\left( \frac{c_1e^{\beta (x+\frac{7\beta t}3)}\beta }{%
3c_1e^{\beta (x+\frac{7\beta t}3)}-\frac{3\alpha }\beta }\right) ^{\frac 23}.
\label{22}
\end{equation}

3. The fifth order the NW(KPP) equation is discussed as follow:

\begin{equation}
u_t-u_{xx}+au+cu^3+eu^5=0.  \label{23}
\end{equation}
A transformation for equation(23) is introduced

\begin{equation}
u(x,t)=\frac{g(x,t)}{\sqrt{f(x,t)}}+u_0(x,t),  \label{24}
\end{equation}
substituting the formula(24) into the equation(23) with {\it Mathematica},
make the coefficients of like powers of $f(x,t)$ vanish, so as to obtain the
following set equations,

\begin{equation}
\begin{array}{l}
f^{-\frac 12}:ag+3cgu_0^2+5egu_0^4+g_t-g_{xx}=0, \\ 
f^{-1}:3c+10eu_0^2=0, \\ 
f^{-\frac 32}:2cg^3+20eg^3u_0^2-gf_t+2f_xg_x+gf_{xx}=0, \\ 
f^{-2}:5eg^4u_0=0, \\ 
f^{-\frac 52}:4eg^4-3f_x^2=0.
\end{array}
\label{25}
\end{equation}
and $u_0(x,t)$ satisfied the equation

\begin{equation}
u_{0t}-u_{0xx}+au_0+cu_0^3+eu_0^5=0,  \label{26}
\end{equation}
which has the similar from of the equation (23), thus the set of the
equations (24)-(26) constitutes an auto-B\"{a}cklund transformation. It's
obviously, the equations(26) have a trivial solution $u_0(x,t)=0$. Based on
the trivial solution, introducing the similarity variable $\xi =x-vt$, and
taking the following ansatz

\begin{equation}
f^{\prime }(\xi )=\alpha f(\xi )+\beta f(\xi )^2,  \label{27}
\end{equation}
where $\alpha $ and $\beta $ are undetermined parameters. Substituting (27)
into the equations (25), choosing

\begin{equation}
\begin{array}{l}
v=\frac 13(2\sqrt{3}\sqrt{-4a+\frac{c^2}e}-\sqrt{3}c\sqrt{\frac 1e)}, \\ 
\alpha =\frac 12(-\sqrt{3}c\sqrt{\frac 1e}-v),
\end{array}
\label{28}
\end{equation}
the equations (25) become identical equations, then the equation(27) can be
solved just by integration. So the solution of the equation (23) is

\begin{equation}
u(x,t)=(\frac 3{4e})^{\frac 14}\sqrt{\frac{\alpha e^{c_1}}{e^{c_1}-e^{\alpha
(x+vt)}\beta }}.  \label{29}
\end{equation}

In conclusion, a method to construct the exact solution of the PDE is
presented in this paper, which combine the two kind methods(the nonlinear
transformation and RQ method). The 1+1 D nonlinear diffusion equation are
discussed as example to show this method. When the parameter $a_i$ in the
equation(6) take four kinds of value, the four equations are discussed. For
every equation, the auto-B\"{a}cklund transformation of it is presented.

{\bf Acknowledgments: }This work was supported by the National Natural
Science Foundation of China. The corresponding email address:
yangkq@lzu.edu.cn (Yang Kongqing).


\begin{references}
\bibitem{1}  C.S.Gardner, J.M.Greene, M.D.Kruskal, R.M.Miura,
Phys.Rev.Lett., 19(1067), 1095.

\bibitem{2}  M.J.Skrinjar, D.V.Kapor, S.D.Stojanovic, J.Phys.Condens.Matter,
32(1989), 725.

\bibitem{3}  Nozaki.K, J.Phys.Soc.Japan, 56(1987), 3052.

\bibitem{4}  R.Hirota, J.Math.Phys, 14(1973), 810.

\bibitem{5}  J.Weiss, M.Tabor, G.Carnevale, J.Math.Phys, 24 (1983), 522.

\bibitem{6}  F.Cariello, M.Tabor, Physica D,53 (1991), 59.

\bibitem{7}  M.Wang, Phys.Lett.A, 216 (1995), 67.

\bibitem{8}  L.Yang, Z.Zhu,Y.Wang, Phys.Lett.A, 260 (1999), 55

\bibitem{9}  M.Otwinowski, al, Phys.Lett.A, 128 (1988), 483.

\bibitem{10}  L.Huibin, W.Kelin, J.Phys.Math.Gen, 23(1990), 3923.
\end{references}
\end{document}